\documentclass[fleqn,usenatbib,useAMS]{mnras}

\usepackage{newtxtext,newtxmath}

\usepackage[T1]{fontenc}

\usepackage{graphicx}	
\usepackage{amsmath}	
\usepackage{amssymb}	

\newcommand{\fe}{Fe~K$\alpha$}
\newcommand{\etal}{et al.}
\newcommand{\mrk}{Mrk~335}

\title[Ionisation from a Lamppost X-ray Source]{Ionised Accretion Discs in Active Galactic Nuclei: The Effects of a Lamppost with a Variable Height}

\author[D. R. Ballantyne]{
D. R. Ballantyne\thanks{E-mail: david.ballantyne@physics.gatech.edu}
\\
Center for Relativistic Astrophysics, School of Physics, Georgia
  Institute of Technology, 837 State Street, Atlanta, GA 30332-0430, USA\\
}

\date{Accepted XXX. Received YYY; in original form ZZZ}

\pubyear{2017}

\begin{document}
\label{firstpage}
\pagerange{\pageref{firstpage}--\pageref{lastpage}}
\maketitle

\begin{abstract}
The X-ray emitting corona irradiates and ionises the surface of the
inner accretion disc in Active Galactic Nuclei (AGNs). The ionisation
parameter of the inner disc at a radius $r$ from the black hole,
$\xi(r)$, can be used to infer
information about the location of the corona. Here, a new formula is derived
that predicts $\xi(r,h)$ for a disc irradiated by a X-ray source at a
height $h$ above the black hole symmetry axis (i.e., a lamppost
geometry). The equation is independent of the black hole mass and the
X-ray spectrum, and accounts for the effects of gravitational
light-bending on the ionisation state and a variable coronal
dissipation factor. We predict a strong ionisation gradient across
the surface of the inner disc that depends on the black hole spin and
lamppost height. For a fixed $h$, the ionisation parameter is also
expected to increase as $\lambda^3$, where $\lambda$ is the observed
bolometric Eddington ratio of the AGN. Comparing this formula to
the observed $\xi$-$\lambda$ relationship for \mrk\ yields $h \propto
\lambda^{0.5-0.6}$, supporting the view of a dynamic X-ray corona in AGNs. 
\end{abstract}




\section{Introduction}
\label{sect:intro}
X-ray spectroscopy is an invaluable probe of the physics of accreting
supermassive black holes in active galactic nuclei (AGNs). The X-ray spectra of AGNs are largely comprised of
two components: a cutoff power-law and a
reflection spectrum that is produced by the reprocessing of the
power-law in dense matter \citep[e.g.,][]{fr10}. The observed relativistically broadened
\fe\ lines indicate that the inner regions of the accretion disc are
responsible for at least some of this reprocessing. Thus, as supported by its rapid variability \citep[e.g.,][]{kara16}, the illuminating power-law source must also arise from small
radii, close to the central black hole. The power-law is
consistent with arising from the Compton up-scattering of ultraviolet
photons from the accretion disc in a tenuous and hot electron corona
\citep[e.g.,][]{hm91,hm93}. Therefore, the reflection signal and coronal emission are
deeply intertwined within the total spectrum, and one of the
challenges of X-ray observations of AGNs is to extract the physical
conditions of the disc-corona system from the data.

One property of the
accretion disc that can yield important insights is the ionisation
state produced by the irradiating power-law. The surface ionisation
state is sensitive to both the disc structure and the coronal
illumination --- most simply described by an ionisation parameter,
$\xi = 4 \pi F_x/n_{\mathrm{H}}$, where $F_x$ is the X-ray flux
incident on gas with hydrogen number density
$n_{\mathrm{H}}$. Calculations of X-ray reprocessing from the surface
of accretion discs show that changes in $\xi$ can have a significant
and measurable effect on the observed reflection spectrum
\citep[e.g.,][]{garcia13,garcia14}. Therefore, measurements of $\xi$ from AGNs have the potential
to probe the illumination and structure of the disc-corona system
close to the black hole.

Interpretation of any measured values of $\xi$ are hampered by
uncertainties in the location and geometry of the corona, as well as
the exact physical structure of accretion discs at distances $r \la
10$~$r_g$ from the black hole ($r_g=GM/c^2$ is the gravitational
radius of a black hole of mass $M$). As a first step, \citet*{bmr11}
used a \citet{ss73} based $\alpha$-disc model to derive an analytical
estimate of how $\xi(r)$ should depend on various interesting physical
parameters such as black hole spin ($a$), coronal dissipation fraction
($f$; \citealt{sz94}), and Eddington ratio
($\lambda=L_{\mathrm{bol}}/L_{\mathrm{Edd}}$, where $L_{\mathrm{bol}}$
is the bolometric luminosity and $L_{\mathrm{Edd}}=4\pi GM
m_{p}c/\sigma_{\mathrm{T}}$ is the Eddington luminosity). The
calculation predicted a strong correlation between $\xi(r)$ and
$\lambda$ which was confirmed using a selection of literature values,
although the data indicated a flatter slope than the prediction. 

The derivation by \citet{bmr11} assumed a compact, geometrically thick
corona (where the coronal scale-height $h = r$) and neglected
relativistic effects that could enhance the irradiating flux. In recent years, \fe\ reverberation measurements and quasar
micro-lensing studies point to a coronal geometry that is very compact
($r \la 10$~$r_g$), and may be better approximated by a lamppost geometry where the X-ray source is situated above the spin
axis of the black hole \citep{rm13,kara16}. Here, we re-derive the expected
$\xi(r)$ relationship for a radiation-pressure dominated accretion
disc irradiated by a lamppost X-ray source, and include the effects of
relativistic light-bending. As is shown below, with the appropriate
datasets the predicted $\xi(r,h)$ relationship can be used to determine
the potential evolution of the X-ray emitting corona. The next section
describes the derivation of the new $\xi(r,h)$ relationship, with the
results analyzed in Sect.~\ref{sect:results}, including an application
to the AGN Mrk~335. A summary and
conclusions are presented in Sect.~\ref{sect:concl}.

\section{Calculations}
\label{sect:calc}
As in \citet{bmr11} we begin by writing $\xi=4\pi m_{p} \rho^{-1}
F_x$, where $m_p$ is the mass of a proton and $\rho$ is the density of
a radiation-pressure dominated $\alpha$ disc \citep{ss73}, corrected for a
non-zero coronal dissipation fraction\footnote{This is a fixed
  fraction of the total dissipation profile, which is centrally
  concentrated \citep{ss73}, and is a decent approximation to a
  lamppost corona.} by following \citet{sz94}:
\begin{equation}
\begin{aligned}
\rho = & (2.23\times 10^{-6}) \left ( {\eta \over 0.1} \right )^2 \left
( {\alpha \over 0.1} \right )^{-1} \left ( {M \over M_{\odot}} \right
)^{-1} \lambda^{-2} \left ( {r \over r_g} \right )^{3/2} \\ 
 & \times R_z^2 R_T R_R^{-3} (1-f)^{-3}\ \mathrm{g\ cm^{-3}}.
\label{eq:density}
\end{aligned}
\end{equation}
In this equation, $\eta$ is the radiative efficiency of the accretion
process, and $(R_R, R_z, R_T)$ are relativistic corrections to the Newtonian
$\alpha$-disc equations that depend on $a$ and $r$ \citep{kro99}. We
assume this is the density responsible for X-ray reflection at the
disc surface.

The problem is now to determine the total X-ray flux $F_x$ incident on the surface of
the disc at a radius $r$ from a X-ray source radiating at a height $h$
(in units of $r_g$) above the black hole while including the
relativistic focusing effects. Following \citet{vincent16}, we define
a lamppost X-ray source emitting isotropically in its rest-frame with a
spectrum $\nu_{\mathrm{src}}^{-\beta}$. The flux normal to the disc
surface at some radius $r$ is then 
\begin{equation}
F_{X,\nu_{\mathrm{disc}}}(r)=A \nu_{\mathrm{disc}}^{-\beta}
\mathcal{F}(r,h),
\label{eq:f1}
\end{equation}
where $A$ is a normalization constant, and $\mathcal{F}(r,h)$ is a function that
describes the illumination profile of the disc irradiated by a
lamppost source and includes the
effects of light-bending. The measured frequency of a photon striking the disc,
$\nu_{\mathrm{disc}}$ is related to its frequency at the source
$\nu_{\mathrm{src}}$ in a lamppost geometry by \citep{dauser13}
\begin{equation}
\label{eq:glp}
g_{\mathrm{lp}}={\nu_{\mathrm{disc}} \over \nu_{\mathrm{src}} } =
{r^{3/2} + a \over \sqrt{r^3+2ar^{3/2}-3r^2}} \sqrt{{h^2 + a^2 -2h}
  \over h^2 +a^2}.
\end{equation}

The function $\mathcal{F}(r,h)$ has been computed using
ray-tracing simulations by several groups \citep[e.g.,][]{fk07,cy12,wf12,dauser13,vincent16}. Writing
$\mathcal{F}(r,h)=F(r,h)g_{\mathrm{lp}}^{1+\beta}$, \citet{fk07} provide
fitting formulas for the shape of the irradiation pattern $F(r,h)$ for
$3 \leq (h/r_g) \leq 100$ and $r > 1.15$~$r_g$ (their Eqs.~10 and 11). With this
identification, Eq.~\ref{eq:f1} can be integrated over frequency to
yield the total X-ray flux normally incident on the disc at radius
$r$:
\begin{equation}
F_X=AF(r,h)g_{\mathrm{lp}}^{1+\beta}(\nu_{\mathrm{disc,max}}^{1-\beta}-\nu_{\mathrm{disc,min}}^{1-\beta})(1-\beta)^{-1}.
\label{eq:f2}
\end{equation}

To set the normalization $A$, we make use of the X-ray
luminosity of the lamppost as seen by the accretion disc, and can be defined by integrating
the flux (Eq.~\ref{eq:f2}) over the entire accretion disc, and
converting the frequencies to their values in the source frame (using
Eq.~\ref{eq:glp}):
\begin{equation}
L_X=A(\nu_{\mathrm{src,max}}^{1-\beta}-\nu_{\mathrm{src,min}}^{1-\beta})(1-\beta)^{-1}
\mathcal{A},
\label{eq:lx}
\end{equation}
where
\begin{equation}
\mathcal{A}=\int_{r_{\mathrm{in}}}^{r_{\mathrm{out}}} F(r,h)
g_{\mathrm{lp}}^2 dS(r),
\label{eq:areaint}
\end{equation}
and
\begin{equation}
dS(r)=2\pi r\sqrt{{r^2 + a^2 + 2a^2/r} \over {r^2-2r+a^2}} dr
\label{eq:ds}
\end{equation}
is the proper element of disc area at the midplane \citep{vincent16}. Equations~\ref{eq:f2} and \ref{eq:lx} can be used to eliminate
$A$ and obtain a final expression for the X-ray flux incident on the
disc at radius $r$,
\begin{equation}
F_X(r)={L_X F(r,h)g_\mathrm{lp}^2 \over z(M)\mathcal{A}},
\label{eq:f3}
\end{equation}
where
\begin{equation}
z(M)=\left ( {GM_{\odot} \over c^2} \right )^2 \left ( {M \over
    M_{\odot}} \right )^2
\label{eq:zed}
\end{equation}
is a factor that converts the area to physical units (e.g., cm$^2$).

As mentioned above, $L_X$ is the total X-ray luminosity seen by the
disc, and will be directly related to the accretion power dissipated
in the corona, $L_X=fL_{\mathrm{bol}}$.
With this identification, the expression for the flux on the disc
surface (Eq.~\ref{eq:f3}) can be combined with the one for the disc
density (Eq.~\ref{eq:density}), to yield the desired equation for the disc ionisation
parameter at radius $r$ when illuminated by a lamppost corona at
height $h$ while including the effects of gravitational light-bending:
\begin{equation}
\begin{aligned}
\xi(r,h)= & (5.44\times 10^{10}) \left ({\eta \over 0.1} \right )^{-2} \left (
  {\alpha \over 0.1} \right) \lambda^3 \left ( {r \over r_g} \right )^{-3/2} R_z^{-2} R_T^{-1} \\
  & \times R_R^3 f(1-f)^3 F(r,h) g_{lp}^2 \mathcal{A}^{-1} \mathrm{erg\ s\ cm^{-1}}.
\label{eq:newxi}
\end{aligned}
\end{equation}
Note that $\xi$ only depends on the black hole spin and not on its
mass or the spectral shape of the illuminating X-rays. The equation is valid for $3\ r_g \leq h \leq 100$~$r_g$ and $r > 1.15$~$r_g$, assuming $a$
is large enough to ensure the disc extends to such small radii. In
what follows, $\mathcal{A}$ is calculated with
$r_{\mathrm{out}}=200$~$r_g$ and $r_{\mathrm{in}}$ is set to the
innermost stable circular orbit (ISCO).

\section{Results}
\label{sect:results}
\begin{figure}
\includegraphics[angle=-90,width=0.48\textwidth]{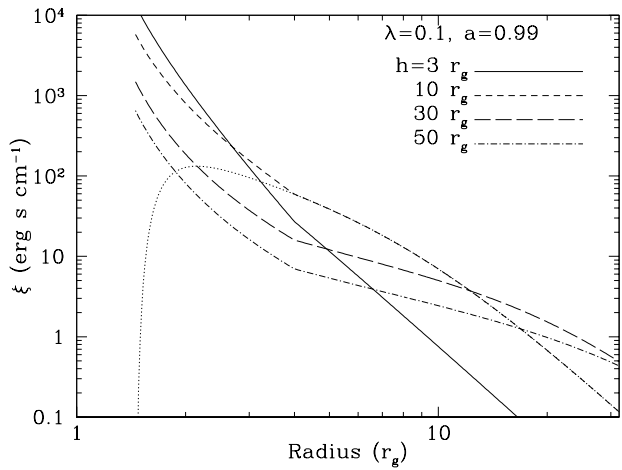}
\caption{The predicted $\xi(r)$ of an irradiated accretion
disc from a lamppost X-ray source situated at different heights above
the black hole (Eq.~\ref{eq:newxi}). For this illustration,
$\lambda=0.1$, $a=0.99$, $f=0.45$ \citep{vf07}, $\alpha=0.1$ and
$\eta=0.1$. The different line styles indicate the various source
heights considered. The ($R_z$,$R_T$,$R_R$) factors
are fixed at their $r=4$~$r_g$ values for $r < 4$~$r_g$ to avoid an
unphysical turnover in $\xi$. The dotted line shows an example of this
behavior for $h=10$~$r_g$.}
\label{fig:xivsr}
\end{figure}

Figure~\ref{fig:xivsr}
plots $\xi(r,h)$ across the inner accretion disc
predicted by Eq.~\ref{eq:newxi} for lampposts with 4 different heights. For the purpose of
this example, a fixed coronal dissipation fraction of $f=0.45$ \citep{vf07}
and Eddington ratio $\lambda=0.1$ are used for the
calculations. Unlike earlier calculations in the
literature \citep[e.g.,][]{svo12,cy12,cy15}, these ionisation profiles
include the effects of both relativistic light-bending and the density
profile of an accretion disc model. The combination of the radial
dependicies of both the disc illumination and density leads to the
predicted shape of the ionisation profiles.

Apart from the one shown using the dotted line, each of the ionisation
profiles shown in Fig.~\ref{fig:xivsr} exhibits a break at
$r= 4$~$r_g$. The break results from setting the ($R_z$,$R_T$,$R_R$)
factors at $r < 4$~$r_g$ to their values at $r=4$~$r_g$
to correct for the divergence of $R_R$ and $R_T$ \citep{kro99} as the radius
approaches the ISCO. As $r$
decreases past $4$~$r_g$, $R_R$ and $R_T$
plunge toward zero, causing the density to become infinitely large
(Eq.~\ref{eq:density}) and hence $\xi(r,h)$ approaches zero at the
inner edge of the disc (dotted line in Fig.~\ref{fig:xivsr}; see also \citealt{kb16}, Fig. 6). However, as
shown by numerical simulations of accretion flows, this behaviour is
unphysical as the disc density actually drops as the gas passes through the
ISCO \citep[e.g.,][]{teix14}, and the ionisation parameter should therefore increase at
small radii \citep{rf08}. After fixing the factors, the predicted ionisation parameters increases
strongly to small radii, and more closely follows the increase in
illumination predicted by light-bending models \citep[e.g.,][]{dauser13}; however, there is
still some effect from the radial dependence of the density. Thus, our correction to the relativistic factors
should give a more accurate portrayal of $\xi(r)$ close to the
ISCO. Ultimately, numerical simulations of the ionisation profile
along the disc surface are needed to precisely determine $\xi(r)$.

\begin{figure}
\includegraphics[angle=-90,width=0.48\textwidth]{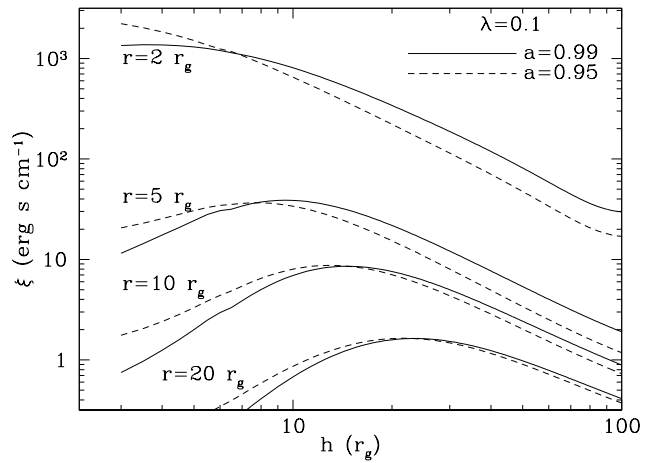}
\caption{The variation of $\xi$ with lamppost height $h$ at four radii
  along the accretion disc. The model parameters are the same as
  described in Fig.~\ref{fig:xivsr}. The ionisation parameter close to
  the ISCO falls as the source height increases. However, because of
  the reduction of gravitational focusing to small radii, $\xi(r,h)$ at $r \ga 5$~$r_g$ actually rises
  as $h$ increases before reaching
  a maximum and then dropping. The dashed lines show results for a model
with a slightly lower black hole spin ($a=0.95$) where the larger
ISCO means that more flux can be distributed over the disc leading to
slightly elevated values of $\xi$ for low $h$ compared to the high
spin model.} 
\label{fig:xivsh}
\end{figure}

Figure~\ref{fig:xivsr} clearly shows the impact of gravitational lightbending on
the ionisation state of the inner disc. For a source height of
$3$~$r_g$, most of the released luminosity of the lamppost is focused
onto the disc close to the ISCO, with only a small amount of flux
irradiating the disc beyond $\sim 10$~$r_g$. As the lamppost is raised
higher, the amount of flux reaching larger radii increases, giving rise to the
seemingly paradoxical condition where the source appears brighter to
the disc even though it is moving further away
(Figure~\ref{fig:xivsh}). 

\begin{figure}
\includegraphics[angle=-90,width=0.48\textwidth]{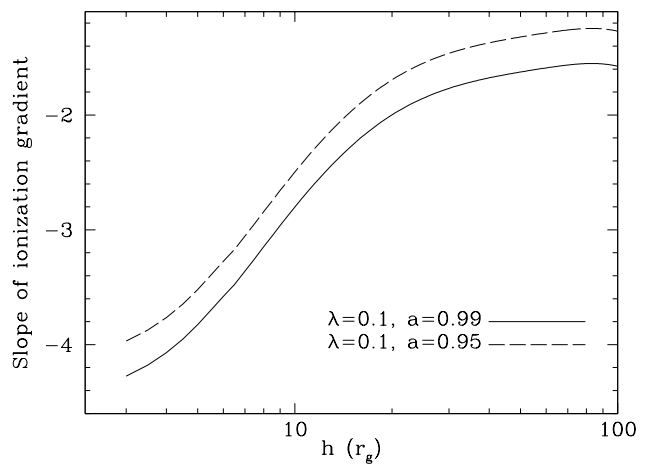}
\caption{The slope of the disc surface ionisation gradient, defined as
  $\log \xi \propto b\log (r/r_g)$, predicted by Eq.~\ref{eq:newxi} as
  a function of the lamppost height. The slope, $b$, is computed
  between $r=2$ and $20$~$r_g$ for two different black hole spins. The
  slope of the ionisation gradient increase with the spin because the
  smaller ISCO allows more radiation to be focused to the center of
  the disc. Values for $\alpha$,
  $\eta$ and $f$ are fixed to $0.1$, $0.1$, and $0.45$, respectively.}
\label{fig:xigradient}
\end{figure}

A significant ionisation gradient on the surface of the
inner accretion disc is predicted, with a slope that depends strongly on the height
of the lamppost. This gradient is a combination of both the
illumination pattern along the disc (often called the disc emissivity
when computing the relativistic effects on the reflected spectrum;
\citealt{dauser13}), and the radial density structure of the disc. To
quantify the gradient in the models shown in Fig.~\ref{fig:xivsr}, the
slope $b$ of the relation $\log \xi \propto b \log(r/r_g)$ was
computed over $2 \leq (r/r_g) \leq 20$ as a function of source height.
Figure~\ref{fig:xigradient} shows the resulting ionisation gradient
varies from $\xi \propto (r/r_g)^{-4.3}$ for $h=3$~$r_g$ to $\xi
\propto (r/r_g)^{-1.6}$ for $h \ga 50$~$r_g$.  
This result may assist the interpretation of ionisation parameters
determined from reflection models fit to AGN X-ray data \citep[e.g.,][]{br09}, since
the measured value of $\xi$ is actually an average over an unknown ionisation
gradient. As the disc emissivity is included in the calculations,
Fig.~\ref{fig:xigradient} shows that for a lamppost source the
observed $\xi$ will be dominated by the ionisation state of the inner
accretion disc.

The principle impact of a different black hole spin on $\xi(r,h)$ is the change
in the location of the ISCO. This affects how the flux can be
distributed along the disc \citep[e.g.,][]{fk07} and is illustrated in
Figs.~\ref{fig:xivsh} and~\ref{fig:xigradient}. In these plots, the
results for $a=0.95$ ($r_{\mathrm{ISCO}}=1.94$~$r_g$) and $a=0.99$
($r_{\mathrm{ISCO}}=1.45$~$r_g$) are shown. The disc extends closer to
the black hole in the $a=0.99$ case and therefore gravitational
light-bending can strongly focus flux to small radii leading to a
large $\xi$ at the ISCO and a steep
gradient. Therefore, when comparing $\xi$ at the same radius, the lower
spin model has the larger $\xi$ when $h$ is small because more flux
impacts these radii. When $h$ is larger, the
situation reverses, as the high spin model allows for more photon
trajectories to impact the inner disc.

\begin{figure}
\includegraphics[angle=-90,width=0.48\textwidth]{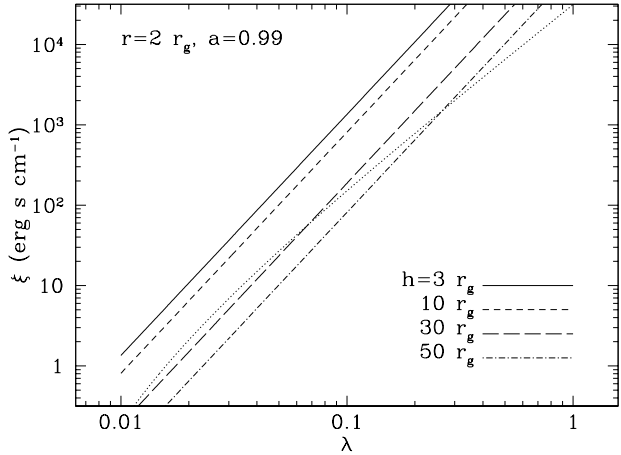}
\caption{The predicted $\xi$-$\lambda$ relationship at $r=2$~$r_g$ for
  different lamppost heights
  (Eq.~\ref{eq:newxi}). All lines, except for the dotted line, plot
  models where $f=0.45$, $\eta=0.1$ and $\alpha=0.1$. The dotted line
  shows how the $h=30$~$r_g$ model changes when $\log f \propto
  -0.77\log \lambda$ \citep{sr84}.}
\label{fig:xivsratiowh}
\end{figure}

Finally, Figure~\ref{fig:xivsratiowh} shows the strong dependence that
$\xi(r,h)$ is predicted to have on $\lambda$.
Equation~\ref{eq:newxi} predicts that $\xi \propto \lambda^3$,
which is the same dependence as the \citet{bmr11} relationship. As
shown in that paper, this relationship can be altered if some of the
parameters of Eq.~\ref{eq:newxi} also depend on $\lambda$. For
example, the coronal dissipation is likely a function of the Eddington
ratio, and the dotted line in Fig.~\ref{fig:xivsratiowh} shows how the
predicted $\xi$-$\lambda$ relationship changes if $\log f \propto
-0.77\log \lambda$ \citep{sr84}. Of course, this new prediction of $\xi$
depends significantly on lamppost height ($h$). Therefore, if the
height of the X-ray source changes with $\lambda$ \citep{kara16}, then the
predicted $\xi$-$\lambda$ relationship will be affected. Finally,
Fig.~\ref{fig:xivsratiowh} shows that the predicted ionisation
parameter of the inner disc can be very low for Eddington ratios of a
few percent -- typical values for Seyfert galaxies. Indeed, the model
naturally explains why most relativistic lines from Seyfert galaxies
arise from neutral iron \citep[e.g.,][]{br09} and
why rapidly accreting AGNs, such as Mrk 335 (discussed below) or
Ark~564 \citep{kara17} exhibit lines from ionised iron (see
also the stacking results of \citealt{liu16}).

\subsection{Application to Mrk 335}
\label{sub:mrk335}
As a specific example of how Eq.~\ref{eq:newxi} can be used to
elucidate details of AGN coronae, we consider the case of \mrk. This AGN was the
subject of a recent study by \citet{kb16} that self-consistently
modeled a dozen separate X-ray observations with relativistic ionised
reflection models. The observations spanned nearly a decade in flux
and were collected over a period of 14 years. Thus, they provided many
individual snapshots of the state of the inner accretion disc when the
source was accreting at different rates. The analysis of \citet{kb16}
found a strong correlation between the measured $\xi$ and
the observed flux which, after estimating bolometric corrections, gave
a correlation with $\lambda$. 

\begin{figure}
\includegraphics[angle=-90,width=0.48\textwidth]{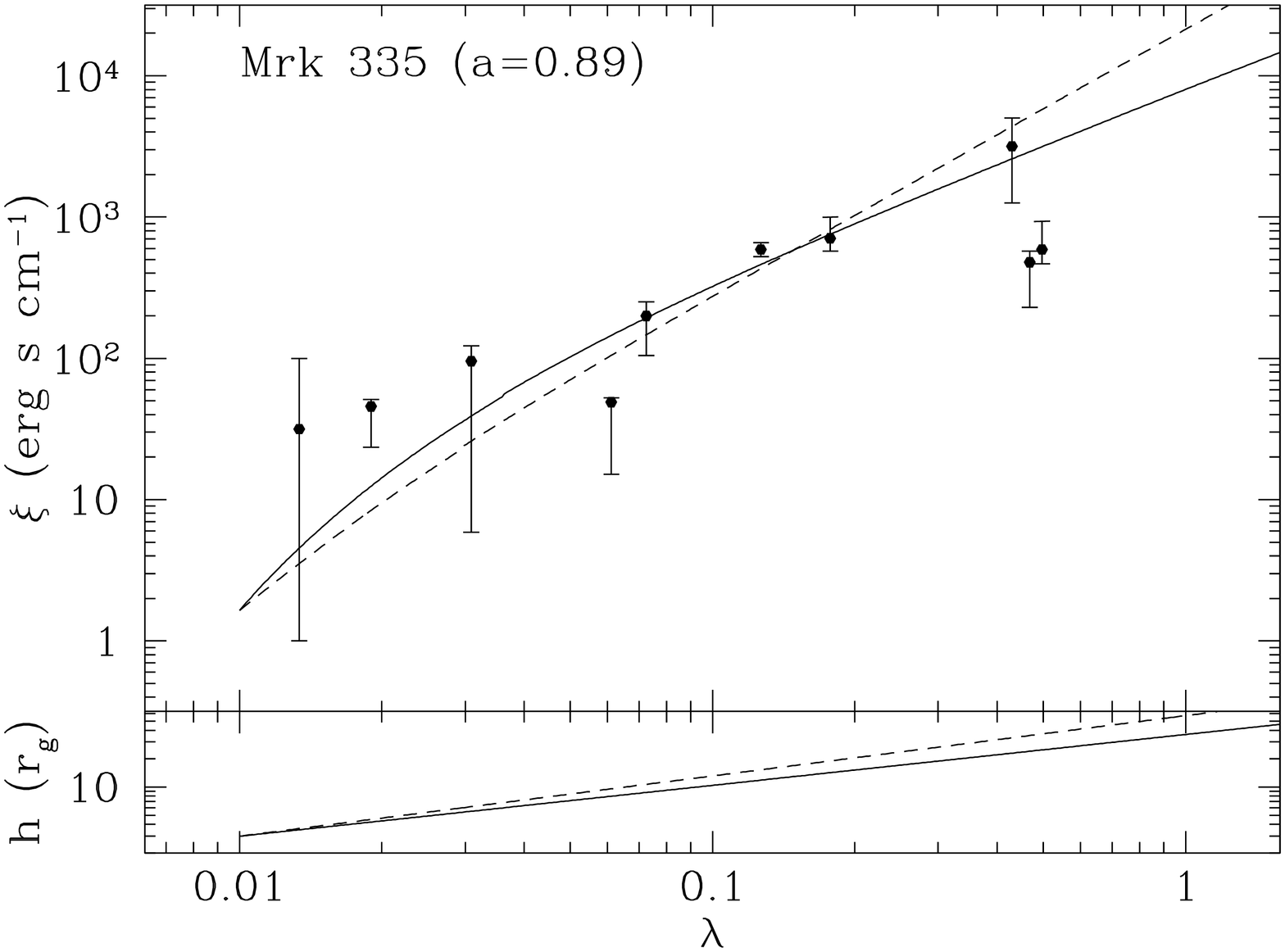}
\caption{The top panel plots the $\xi$ versus $\lambda$ data derived
  from spectral fitting of multiple observations of the AGN
  \mrk\ \citep{kb16}. The two lines are models fit to these data using
Eq.~\ref{eq:newxi} under the assumption $\log h \propto \beta \log
\lambda$. The solid line plots a model where $\log f \propto
  -0.77\log \lambda$ \citep{sr84} and the dashed line assumes $f=0.45$
  \citep{vf07}. Both models are calculated assuming the observed $\xi$ is
  from the ISCO ($r\approx 2.4$~$r_g$) for a black hole with spin
  $a=0.89$ \citep{kb16}. In addition, $\eta=0.1$, $\alpha=0.3$, and
$h=3$~$r_g$ for $\lambda=0.01$. The lower panel shows how $h$ varies
  with $\lambda$ for the two models: $h \propto \lambda^{0.54}$ (solid
line) and $h\propto \lambda^{0.64}$ (dashed line).}
\label{fig:mrk335}
\end{figure}

The upper panel of Figure~\ref{fig:mrk335} shows the measured \mrk\
$\xi$-$\lambda$ correlation from \citet{kb16}.
Overplotted in this panel are two models calculated directly from
Eq.~\ref{eq:newxi} and are described in detail below. The comparison
between the data and the models is not entirely self-consistent. The
quality of the \mrk\ spectra meant that the disc emissivity needed to
be fixed at $-3$ during spectral fitting. The values of $\xi$ computed
from Eq.~\ref{eq:newxi} incorporate the correct emissivity for the
lamppost model, which will vary with the height of the X-ray source
\citep{dauser13}. Because of this inconsistency, the application to
\mrk\ presented here should be considered an illustration of the
potential of this analysis for future datasets where the emissivity
can be satisfactorily constrained.

To calculate the expected $\xi$-$\lambda$ relations, $a$ was set to
$0.89$, the value determined by the spectral fitting of
\citet{kb16}. The measured values of $\xi$ were derived by \citet{kb16} assuming reflection down to
the ISCO with a fixed emissivity of $-3$. Therefore, the measured
values of $\xi$ are determined from a centrally concentrated spectral
model, similar to a lamppost geometry. Figure~\ref{fig:xigradient}
shows that the ionisation gradient is expected
to be steep, and the predicted
ionisation parameter from the ISCO should dominate the reflected
emission for this geometry. To compare against the data, the two $\xi$-$\lambda$ models shown in
Fig.~\ref{fig:mrk335} are therefore computed at the ISCO of an accretion
disc around a black hole with spin $a=0.89$ (i.e.,
$r=2.4$~$r_g$; both models also assume $\eta=0.1$). The dashed line assumes a
constant $f=0.45$ \citep{vf07} and the solid line shows a model where $\log f
\propto -0.77\log \lambda$ \citep{sr84}. However, in order to account for the slope of the observed
relation ($\log \xi \propto 1.1\log \lambda$; \citealt{kb16}), the
height of the lamppost must also vary with Eddington ratio. 

To determine the variation of $h$ with $\lambda$ we assume $\log h \propto \beta \log \lambda$. To
normalize the relation, $h$ is set to 3~$r_g$ (the lowest acceptable
value for Eq.~\ref{eq:newxi}) at $\lambda=0.01$. Even with the X-ray
source at such a low height, the predicted $\xi$ at $\lambda \approx
0.01$ is well below the observed data, so $\alpha$ was
set to $0.3$, consistent
with values found close to the ISCO in numerical simulations of accretion flows
\citep{penna13}. Chi-square fitting was then performed to find the best $\beta$
for each of the two models. For the model with $f=0.45$ (dashed line),
$h\propto \lambda^{0.64}$, while for the model with $\log f \propto
-0.77\log \lambda$ (solid line), $h \propto \lambda^{0.54}$. In both
cases, the lower panel of Fig.~\ref{fig:mrk335} shows how the height
of the lamppost depends on the $\lambda$, reaching heights of
$h \sim 30$~$r_g$ when $\lambda \sim 1$. Lamppost heights $\la
30$~$r_g$ are consistent with the coronal distances inferred from
X-ray reverberation \citep*{kara16,klk17}, especially after accounting
for the effects of dilution \citep[e.g.,][]{cack14,cy15} that can increase
the measured distances by a factor of a few.

An increasing X-ray source height with $\lambda$ in \mrk\ is
consistent with other relationships uncovered in the \citet{kb16}
spectral analysis, such as a smaller reflection fraction at larger
$\lambda$. In particular, by using the measured properties of the
X-ray power-law, \citet{kb16} estimated that the coronal
temperature was much higher, and the optical depth lower, at large
$\lambda$. Combining all these relationships
together supports a corona that was compact and relatively cool at low
$\lambda$, but is hotter and more extended at high Eddington
ratios. The above analysis of the $\xi$-$\lambda$ plane supports this
picture, which is also in broad agreement with the results of
\citet{wf15}, who modeled the emissivity profile of the \fe\ line to infer a
corona in \mrk\ that evolved in both the vertical and radial directions. 
  
\section{Summary and Conclusions}
\label{sect:concl}
In this work, a formula for the ionisation parameter of the surface of
an irradiated AGN accretion disc, $\xi(r,h)$, has been derived for a
X-ray source placed at a height $h$ above the symmetry axis of the
black hole (Eq.~\ref{eq:newxi}). This equation takes into account the effects of
gravitational light-bending and focusing of radiation onto the disc
\citep[e.g.,][]{vincent16}, and is independent of black hole mass
and the spectrum produced by the X-ray source. Therefore, it can be
used to test models of AGN coronae within a lamppost geometry.

As shown in Figs.~1--3, the theory predicts a strong
ionisation gradient across the disc surface that depends on $h$ and
the spin of the black hole, $a$. A more rapidly spinning black hole
has a smaller ISCO, and therefore ionised reflection is enhanced close to the
ISCO by gravitational focusing leading to a steeper ionisation gradient for larger values of $a$. As the height of the X-ray source
increases, the reduction of light-bending at small radii first causes $\xi$ to
grow at small radii before eventually dropping with even larger $h$ (Fig.~\ref{fig:xivsh}). 

The $\xi(r,h)$ equation derived here predicts a strong
dependence on $\lambda$
(Fig.~\ref{fig:xivsratiowh}). However, this dependence can be modified if the
height of the lamppost varies with $h$. This was illustrated by
comparing the predictions of Eq.~\ref{eq:newxi} with the
$\xi$-$\lambda$ data determined for the AGN
\mrk\ \citep{kb16}. Figure~\ref{fig:mrk335} showed that if $h \propto
\lambda^{0.5-0.6}$ the predicted $\xi$-$\lambda$ relation could broadly
follow the observed datapoints. As future X-ray spectroscopic studies
of AGNs become more precise, comparisons between how $\xi$ varies with
$\lambda$ with the theory presented here could provide constraints on
the evolution of the X-ray source in the lamppost model.

The conclusion from the \mrk\ comparison is that the height of the
corona, at least in the lamppost model, will increase with Eddington
ratio or AGN flux state. This result agrees with the recent work of
\citet{klk17} who found that the
corona height increased with the AGN radio jet power, which may indicate a
connection between the base of the jet and the X-ray source
\citep[e.g.,][]{mn04}. Enhanced radiation pressure from the disc may also cause the
corona to be at a larger heights during higher accretion rates \citep{belo99}. Additional relations based on the measured reflection fraction
also indicate a more distant X-ray source at larger $\lambda$ \citep{kb16,klk17}. Future work will focus on extending
the theory presented here to predict $\xi(r,h)$ for a moving X-ray
source, and to investigate calculating a simultaneous prediction of
the reflection fraction.








\bsp	
\label{lastpage}
\end{document}